# ELECTRON KINETICS IN STANDING AND MOVING STRIATIONS IN ARGON GAS

Dmitry Levko[1]

CFD Research Corporation, Huntsville, Alabama 35806, USA

E-mail: dima.levko@gmail.com

The electron kinetics in moving and standing striations in direct current and radio-frequency discharges is studied. The discharge current is such that the thermalizing electron-electron collisions are negligible and the hydrodynamic description of electron component of plasma is not valid. Therefore, the one-dimensional hybrid model is used which models the electron component by particle method, while the ions are described using the drift-diffusion approximation. It is obtained that the electron transport is non-local in space. The electron energy distribution in both discharges is of non-equilibrium nature which is responsible for the non-linearity of the ionization frequency. However, their dynamics in both discharges differs significantly. Namely, in the direct current discharge the distribution function is strongly modulated by the striations passage, while in the radio-frequency discharge the electron energy relaxation time is such that the electron distribution function does not react on the oscillating electric field but is defined by the effective electric field.

[1]Present address: Esgee Technologies Inc., Austin, Texas 78746, USA

## I.    Introduction

Positive column of glow discharge is unstable with the respect to different **types** of instabilities, e.g. contraction and stratification.[1,2,3,4,5,6]  These instabilities are usually considered as undesirable because they disturb the homogeneity of plasma column.   For instance, for luminescent lamps this means their failure.   Moving and standing striations were observed in both direct current (DC) and radio-frequency (RF) discharges (see, for instance, Refs. 2,7,8 and references therein).   There are several type of striations which can be described by using either the hydrodynamic or kinetic approach.[9]   At high currents, the electrons are thermalized due to frequent electron-electron Coulomb collisions.   Then, the striations dynamics can be described by the hydrodynamic approach.[8]   At high gas pressures, the hydrodynamic approach is justified by the frequent electron-neutral collisions.   At intermediate pressures, the electron kinetics is non-local and stratification can be described only at the kinetic level.[4]

In Ref. 8, the nature of standing striations in RF discharge in argon was analyzed using the fluid model.   The modeling results were compared with the experimental results and demonstrated rather good agreement.   It was showed that the instability is due to the non-linear dependence of the ionization rate on the electron density.   It was suggested that this non-linearity is due to the stepwise ionization and the Maxwellization of the electron distribution function due to the Coulomb collisions.   Based on this assumption, it was also concluded that the nature of standing striations in RF discharges is similar with the nature of moving striations in DC discharges.[7]   However, the electron Maxwellization is only possible at high discharge currents when there are frequent electron-electron (*e-e*) collisions.[10]   At low currents, these processes are negligible and the discharge stratification can be described only using the kinetic models.

There are several papers presenting comprehensive computational analysis of the striations at high pressure conditions in both unmagnetized[11,12] and magnetized plasmas.[13] To our knowledge, there are no self-consistent kinetic modeling studies of the low-pressure low-current moving and, especially, standing striations although there are numerous theoretical analyses of the electron kinetics in moving striations.[14,15,16,17,18]

In the present paper, the one-dimensional self-consistent hybrid model is used to understand the nature of striations in RF and DC discharges for the conditions when the *e-e* collisions are negligible. In this model, the electrons are modeled at the kinetic level using the Particle-in-Cell/Monte Carlo collisions (PIC/MCC) approach, while the ions are modeled as the fluid.

## II. Numerical model

In order to model the striations, the self-consistent 1D hybrid model developed in Ref. 19 was used. In this model, the electrons are modeled at the kinetic level using the one-dimensional (1D) PIC/MCC method. Four processes were taken into account, namely, three electron-neutral (*e-n*) collisions (elastic, ionization and excitation of the first electronic level of atoms) and *e-e* Coulomb collision. The probability of the *e-n* and *e-e* collision was calculated by

$$P = 1 - \exp\left(-\Delta l/\lambda(\varepsilon_e)\right), \tag{1}$$

where $\Delta l$ is the distance which electron propagates during one time step, and $\lambda(\varepsilon_e)$ is the mean free path of electron having energy $\varepsilon_e$:

$$\lambda(\varepsilon_e) = \frac{1}{n_g[\sigma_{el}(\varepsilon_e)+\sigma_{ex}(\varepsilon_e)+\sigma_{ion}(\varepsilon_e)]+n_e\sigma_C(\varepsilon_e)}. \tag{2}$$

Here, $n_g$ and $n_e$ are the gas and electron number densities, respectively, $\sigma_{el}$, $\sigma_{ex}$ and $\sigma_{ion}$ are, respectively, the elastic, excitation and ionization cross sections, and $\sigma_C$ is the *e-e* Coulomb cross section. The cross sections of *e-n* collisions were taken from the Biagi's database.[20] The probability of collision was compared with the random number, $rnd$, from the interval (0;1). If $P < rnd$, the collision occurred. Then, another random number was generated to define the type of collision.

The ion density is found using the drift-diffusion approximation[21]

$$\frac{\partial n_i}{\partial t} + \frac{\partial \Gamma_i}{\partial x} = R_{ion} - \frac{n_i}{\tau}. \tag{3}$$

Here, $n_i$ and $\Gamma_i$ are, respectively, the ion number density and the flux, and $R_{ion}$ is the ionization source term which is obtained from the particle module. Also, $\tau(x) = \frac{\Lambda^2}{D_a(x)}$ is the ambipolar diffusion time in the radial direction, where $\Lambda = R_{tube}/2.4$ is an effective transverse discharge dimension ($R_{tube}$ is the discharge tube radius) and $D_a$ is the ambipolar diffusion coefficient. The latter was calculated as $D_a(x) = \frac{2}{3}\varepsilon_e(x)\mu_i(x)$, where $\mu_i$ is the ion mobility and $\varepsilon_e$ is the average electron energy at any given computational cell. As was obtained in Ref. 8, this term is enough to obtain the positive column of glow discharge using the 1D model. Then, the 1D model gives reasonable agreement with the self-consistent two-dimensional axisymmetric model.

The term $\frac{n_i}{\tau}$ was also taken into account in the PIC module by removing $\frac{N_e \Delta t}{\tau}$ macro-particles every time step $\Delta t$ from the simulation domain ($N_e$ is the number of electron macro-particles in a given cell). The boundary conditions for Eq. (3) are

$$\Gamma_i n = \frac{n_i v_{th}}{4} + \alpha \mu_i n_i E n. \tag{4}$$

Here, $v_{th}$ is the ion thermal velocity, $E$ is the electric field at the electrode, $n$ is the normal vector to the electrode, and $\alpha = 1$ if $En > 0$, or $\alpha = 0$ otherwise.

To summarize, the model consists of the following sequence of steps performed for each computational time step $\Delta t$:

1)  Check the secondary electron emission from the walls, if specified, and inject the electrons into the simulation domain.

2) Propagation of electron computational particles into new positions. Remove the particles which cross the boundaries.

3) Model the *e-n* and *e-i* collisions using the Monte Carlo algorithms.

4) Electron weighting on a spatial numerical grid to define their space charge density.

5) (*fluid part*) Solve the ion drift-diffusion equation (3).

6) Solve the Poisson's equation for new electron and ion densities, and specified electrodes' potentials.

7) Return to step #1.

It is important to note that Eq. (3) does not contain any artificial dependence of the ionization source term $R_{ion}$ on the electron density. This dependence arises naturally due to the kinetic treatment of electrons in the present hybrid model.

The interelectrode gap in the present studies was 14 cm. The right electrode was grounded. The fixed potential of $\varphi_0 = -150$ V was applied to the left electrode for the DC discharge, and the oscillating potential $\varphi(t) = \varphi_0 \sin(2\pi\omega t)$ with $\omega = 13.56$ MHz was applied to this electrode for the RF discharge. The secondary electron emission coefficient was set equal to $\gamma = 0.2$ for the DC discharge, while $\gamma = 0$ for the RF discharge. The background gas was argon (Ar) at the fixed pressure of 50 Pa and at the temperature of 300 K. Also, the discharge tube radius was taken equal to 1.1 cm as in Ref. 8.

### III. Results and discussion

### A. Direct current discharge

Figure 1 shows the ion density spatial profiles obtained at four different times. One can see that the cathode-anode (CA) gap is divided into the sheath, the quasi-neutral region where the density is ~$1.2 \times 10^{15}$ m$^{-3}$ (also called the negative glow[22]), the plasma column where the plasma density is much smaller than in the negative glow ($\sim 10^{14}$ m$^{-3}$), and the anode sheath. The high-voltage plasma sheath accelerates the electrons being emitted from the cathode due to the ion impact to the energies do not exceeding 100 eV. These are the electrons which are mainly responsible for the discharge maintenance.[22] The cathode sheath repels the plasma electrons; thus, they leave the domain only through the anode and through the walls modeled by the term $\frac{n_i}{\tau}$ in Eq. (3). The ions leave the gap mainly through the cathode sheath. The simulation results have shown that the fraction of the wall losses is $\sim 10\%$ in the total flux balance.

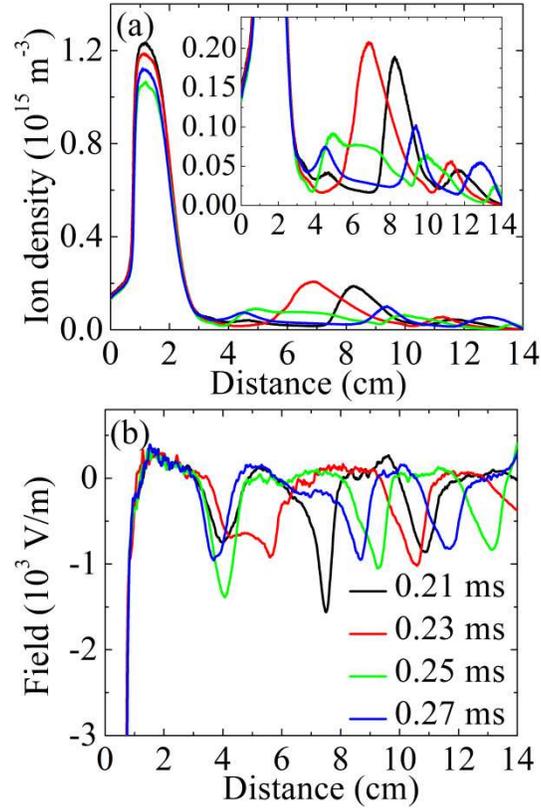

Figure 1. (a) Spatial profiles of the ion density and (b) spatial profiles of the electric field obtained at four different times. Inset in figure (a) is focused on the striations propagating from the anode to the cathode.

The simulations have shown that the largest energy of emitted electrons does not exceed 100 eV, which is much smaller than the cathode potential. On the one hand, this is obtained because the plasma of the positive column is at the positive potential with the respect to the anode ($\varphi_{pl} \sim 20$ V). Therefore, the energy of electrons accelerated in the cathode sheath cannot exceed 130 eV. On the other hand, the cathode sheath is collisional and there is the plasma generation in the sheath. Indeed, in Ar gas, the ionization cross section starts exceeding the momentum transfer one at ~80 eV.[20] The mean free path, $\lambda_e$, of these electrons is ~2 mm, which is shorter than the cathode sheath thickness.

One can conclude from Figure 1(b) that there is the electric field reversal in the negative glow and in several locations of the positive column. In the vicinity of the cathode sheath this is obtained because the energy relaxation length of the secondary emitted electrons, $\lambda_T$, is short and they dissipate large fraction of their energy in the negative glow generating new thermal electrons. The electrons entering the negative glow from the cathode sheath dissipate their energy in inelastic ionizing collisions. Therefore, one can estimate $\lambda_T \approx \lambda_e \sim 2$ mm. One can see that $\lambda_T$ is much shorter than the negative glow length (~ 2 cm, see Figure 1). The slow electrons generated here due to the gas ionization are got trapped in the potential well of ions.

Figure 1(a) shows that the positive column is stratified. One can see 2-3 striations moving from the anode to the cathode. The electric field reversal is obtained in the vicinity of every striation [see Figure 1(b)]. The average velocity of these striations is ~500 m/s, which is comparable with the ion thermal velocity. This means the striations motion occurs on the ion time scale, i.e. the longest time scale in the present model. The electric current to the anode obtained in the present studies is ~3.6 A/cm$^2$. The potential drop within every striation is ~11-13 eV, i.e. slightly smaller than the ionization potential of Ar atom ($\varepsilon_{th} \approx 15.6$ eV). The remaining 3-5 eV necessary for the gas ionization electrons obtain from the electric field presented between two neighboring striations. As was pointed out in Ref. 10, these striations can be adequately described only at the kinetic level because of the non-equilibrium nature of the electron energy distribution. Also, the non-local electron kinetics is typical for these striations.[2]

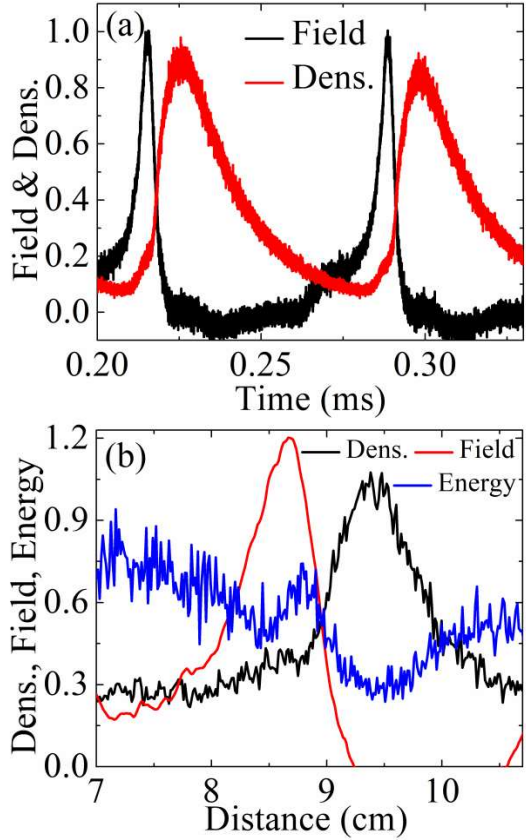

Figure 2. (a) Normalized electric field and plasma density as the functions of time at $x$ = 7 cm, and (b) normalized electron density, electric field and energy obtained at $t$ = 0.27 ms.

Figure 2(a) shows the time dependence of the electric field and the plasma density at $x$ = 7 cm. Similar profiles were obtained in other locations of the positive column. The normalized profiles of the electron density, the electric field and the mean electron energy obtained in the vicinity of striation at $t$ = 0.27 ms are shown in Figure 2(b). One can see the shift between the peaks of the electric field and energy, which means that the electron energy is the non-local function of the electric field.

The motion of striation can be explained as follows.[22] Each of them consists of the neighbor regions of excess of negative and positive space charge.[23] Due to larger mobility, electrons from the negative charge region move toward the positive charge region neutralizing it and leaving behind uncompensated positive charge. This leads to the shift of the maximal value of the electron mean energy with respect to the plasma density [see Figure 2(b)]. Also, the electrons been accelerated in the striation generate quasi-neutral plasma in it. Thus, the region of uncompensated positive charge shifts towards the cathode.

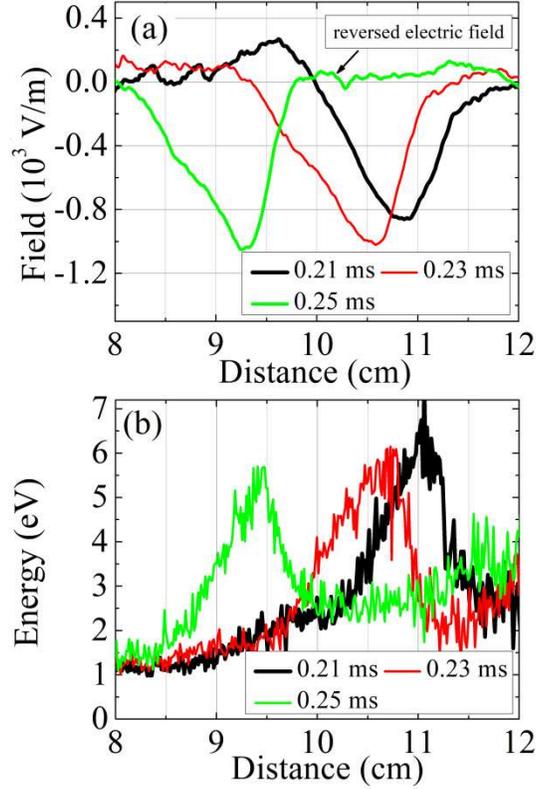

Figure 3. Spatial profiles of (a) the electric field and (b) the average electron energy obtained at three different times.

The plasma density in the positive column is $\sim 10^{14}$ m$^{-3}$. The average electron energy in the plasma column is $\sim$2-4 eV (see Figure 3). The *e-e* Coulomb collision cross section for these electrons is $\sim 2 \times 10^{-17}$ m$^2$, which is $\sim$2 orders of magnitude larger than the *e-n* momentum transfer cross section. However, the electron density is $\sim$8 orders of magnitude smaller than the gas density. Therefore, the influence of *e-e* collisions on the electron energy probability function (EEPF) can be neglected at the given conditions.

Figure 4(a) shows the EEPF obtained at $x \approx 10.5$ cm at three different times. One can see that this distribution is far from the Maxwellian which is due to the insignificant influence of the Coulomb collisions. This means that at the given conditions the Maxwellization of the distribution function cannot be responsible for the non-linearity of the ionization frequency[8] and for the ionization instability onset. Indeed, the additional simulation (not shown here) carried out without accounting for the *e-e* collisions has shown the positive column stratification.

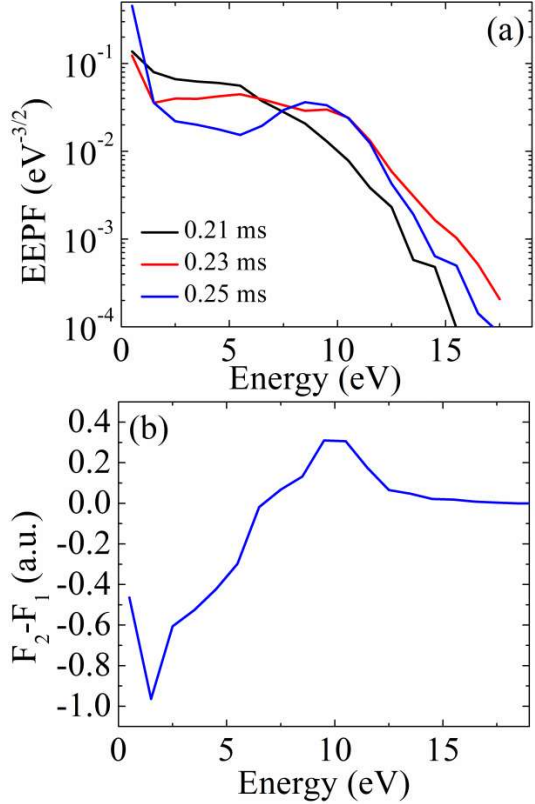

Figure 4. (a) Electron energy probability function obtained at $x \approx 10.5$ cm at three different times, and (b) the difference between distribution functions obtained at $t = 0.23$ ms and $t = 0.21$ ms.

One can conclude from Figure 4(a) that the EEPF shape is modulated by the passage of striations. This is also seen in Figure 4(b) which shows the difference between the EEPFs obtained at $t = 0.23$ ms and $t = 0.21$ ms. The propagation of striations leads to the population/depopulation of the high-energy tail of the distribution function. Similar behavior was measured in Ref. 24 in neon and discussed in Ref. 15 using the theory of the electron bunching.

Figure 4(a) shows that at $t = 0.25$ ms, there is the region of the inversed electron distribution function at 5 eV $< \varepsilon_e <$ 10 eV. Such distributions can lead to the negative absolute conductivity of plasma, if they correspond to the energy range with the sharp increase in the electron momentum transfer cross section.[25] In Ar gas, this is obtained in the range 5 eV $< \varepsilon_e <$ 10 eV, exactly where the inversion of the EEPF is seen in Figure 4(a). Recently, such behavior of the EEPF was analyzed in Ref. 26 in DC discharges in argon and helium. It was shown that the inversion of the EEPF is possible in the regions of glow discharge where the electric field reversal is obtained (e.g., in the negative glow).

The plasma conductivity is defined as $\sigma = \mu_e n_e$, where $\mu_e$ is the electron mobility:[27]

$$\mu_e = -\frac{1}{3n_g}\sqrt{\frac{2q_e}{m_e}}\int_0^\infty \frac{\varepsilon_e}{\sigma_m(\varepsilon_e)}\frac{\partial F(\varepsilon_e)}{\partial \varepsilon_e}d\varepsilon_e. \tag{5}$$

Here, $q_e$ is the elementary charge, $m_e$ is the electron mass, and $F(\varepsilon_e)$ is the EEPF. Integrating this equation over the EEPFs shown in Figure 4(a), one obtains $\sigma \sim 10^{-2}$ $\Omega^{-1}\mathrm{m}^{-1}$, i.e. the plasma conductivity is positive for all distributions. The absolute negative conductivity was not obtained at the conditions of the present studies.

Time evolution of the EEPF can be explained by the following. Striations moving in axial direction lead to the strong temporal and spatial variations of the electric field $E_p(x,t)$ and plasma density $n(x,t)$. The axial electric field $E_p$ consists of the resistive component, $E_r$, and the ambipolar component, $E_a$, which are $E_p = E_r + E_a = \frac{j}{\sigma} + \frac{T_e}{q_e n}\frac{dn}{dx}$.[22] Here, $j$ is the discharge current density, $\sigma$ is the plasma conductivity, $T_e$ is the electron average energy, and $q_e$ is the elementary charge. In this equation, we neglected the spatial gradient of $T_e$ because it is much smaller than the spatial gradient of the plasma density. In the case of moving striations, both terms are non-zero and comparable. Since the density of low-energy part of the EEPF is much larger than the density of the high-energy electrons, the resistive component of the electric field is defined by the low-energy electrons.

Depending on the direction of the plasma density gradient, the direction of $E_a$ may coincide or be opposite to the direction of the resistive component. Namely, at the positive slope of $dn/dx$ the plasma density gradient enhances the plasma field, while at $dn/dx < 0$, the plasma density gradient decreases and even reverses the plasma field [see Figure 1(b) and Figure 2(a)]. The electric field reversal is responsible for the trapping of the low-energy electrons in the vicinity of striations similar to their trapping in the negative glow near the cathode. At the given conditions, the electric field reversal can make negative the electric conductivity of plasma.[26]

The enhancement of the electric field $E_p$ due to the increase of the ambipolar component results in the electron heating to the energies exceeding the ionization threshold of Ar. This leads to the local increase of the plasma density [see Figure 1(a)]. In Refs. 14,16,17 and 18 the process of the electron heating in moving striations was described as the electron bunching in periodic electric field. This is purely kinetic effect which cannot be adequately captured in hydrodynamic models. Note that Figure 4(a) does not show any pronounced electron bunches. The method used in the present studies does not describe the contraction of the EEPF to resonant trajectories. A possible reason for these discrepancies is that the solution of the kinetic equation is carried out over a very large length, which noticeably exceeds the length of the positive column [2].

The hump seen in the energy range 5 eV < $\varepsilon_e$ < 10 eV at $t$ = 0.25 eV is obtained after the passage of striation. It is formed due to the electric field reversal [see Figure 3(a)]. However, the bunching can be obtained after extracting the EEPFs obtained at $t$ = 0.23 ms and at $t$ = 0.21 ms [see the energies ~10 eV in Figure 4(b)]. The population of the high energy tail is also in agreement with the energetic concept presented in Ref. 22. It says that the positive column stratification is "expedient" because the electron has higher probability to get the energy $\varepsilon_{th}$ on the length of one striation than on a few centimeters of unperturbed positive column.

Figure 4 shows that the passage of striation leads to the decrease of the electric field at the given location [see Figure 3(a)] which results in the dissipation of the electron energy in inelastic *e-n* collisions. This leads to the decrease of the electron energy in the tail of the EEPF and to the decrease of the average electron energy [see Figure 3(b)].

### B. Radio-frequency discharge

This subsection presents the results of simulations obtained for the RF discharge (frequency 13.56 MHz). The average discharge current density obtained for these conditions is ~10 A/cm$^2$, which is a few times larger than the current density obtained in DC discharge. This is due to the larger average electron density obtained in the RF discharge.

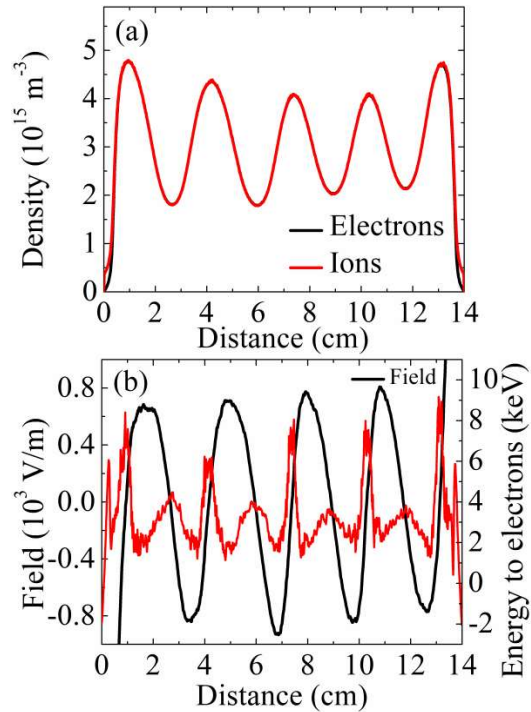

Figure 5. (a) Spatial profiles of the electron and ion densities, and (b) the electron heating power and electric field averaged over the RF period.

Figure 5(a) shows the electron and ion energies averaged over the RF period. One can see the stratification of the plasma column to five unmovable striations. There are two regions in the vicinity of both electrodes which are similar to the negative glow obtained in the DC discharge [compare with Figure 1(a)]. Like the negative glow in DC discharge, these "negative glows" of the RF discharge are also characterized by the electron groups. The slow electrons are the electrons being generated in the vicinity of the sheaths and got trapped in the potential well of ions. The electron trapping also leads to the electric field reversal near the electrodes. The group of fast electrons in the negative glows consists of the electrons which penetrate the sheaths during the sheath collapse and are accelerated during its expansion. These sheaths are collisional and the maximum electron energy obtained at the given conditions near the electrodes does not exceed 80 eV.

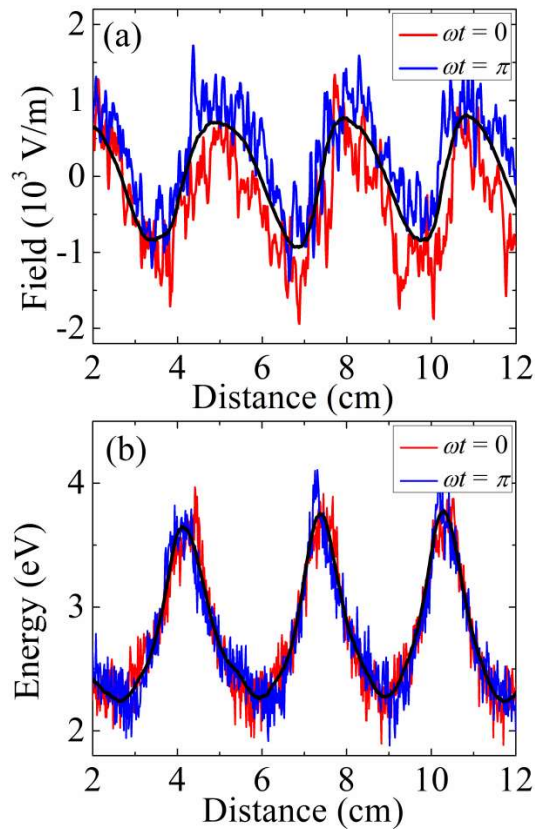

Figure 6. (a) Spatial distribution of the electric field and (b) spatial distribution of the electron average energy obtained at two different times during RF period. Bold black line shows the period-averaged values.

The instantaneous electric field and the average electron temperature at four different times are shown in Figure 6. The bold black lines also show the period-averaged quantities. One can see that the electric field oscillates with the respect to its average value, while the electron average energy remains constant. Like in the case of the DC discharge, the electric field is the sum of the resistive and the ambipolar electric fields, $E_p = \frac{j}{\sigma} + \frac{T_e}{q_e n} \frac{dn}{dx}$. Since the plasma density remains constant far from the sheath, the electric field oscillations are due to the time variation of the conduction current flowing through the plasma. The electron average energy remains constant due to the specifics of the electron kinetics at the given conditions (see discussion below).

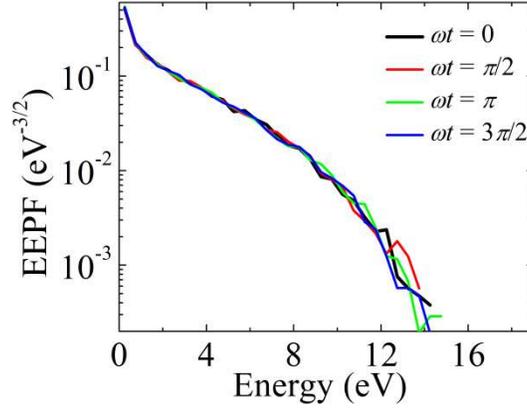

Figure 7. Electron energy probability function obtained near the central striation ($x \approx 6$ cm) at four different times of RF period.

Figure 7 shows the EEPF obtained at four different times of the RF period at $x \approx 6$ cm (in the vicinity of the striation). One can conclude that these functions are far from the Maxwellian distribution which is explained, as in the case of the DC discharge, by the insignificant influence of the *e-e* collisions on its formation.

The largest electron energy seen in Figure 7 is ~20 eV. In this energy range, the EEPF formation is mainly controlled by the *e-n* momentum transfer. This function is formed with a typical time scale $\tau_\varepsilon = 1/(\delta \nu_m)$, where $\delta = 2m_e/m_i$. At the given conditions, one has $\omega \tau_\varepsilon \gg 1$, which means that the electric field varies faster than the energy relaxation occurs.

Then, the EEPF does not depend explicitly on the time varying electric field but is defined by an effective electric field $E_{\text{eff}} = \frac{\tilde{E}}{\sqrt{2}} \frac{\nu_m}{\sqrt{\nu_m^2 + \omega^2}}$,[10] where $\tilde{E}$ is the average electric field. For the gas pressure of 50 Pa and the RF frequency of 13.56 MHz, one obtains that $\nu_m \gg \omega$ for the electron energy in the range 2 eV $< \varepsilon_e <$ 30 eV. Then, the effective field is $E_{\text{eff}} \approx \tilde{E}/\sqrt{2}$.

The kinetics of electrons in stratified RF discharge is of non-local nature. This can be concluded from Figure 5 and Figure 6. They show that the peak of the averaged heating power profile is shifted in space with the respect to the peak of the electric field. The same is obtained for the average electron energy and the plasma density.

The distribution of the electric field is defined by the ratio of the electron energy relaxation length to the striation length.[8] The energy relaxation is due to the momentum transfer collisions and is estimated at the given conditions as ~30 cm. This is much longer than the striation length. This explains the non-local heat transfer between striations.

## IV.   Conclusions

The electron kinetics was analyzed in low-current striations being formed in direct current and radio-frequency discharges. For these striations, the thermalizing electron-electron collisions could be neglected and the electron energy distribution was controlled by the electron-neutral collisions.

It was obtained that the electron kinetics is of non-local nature in both types of discharge. The nonlinearity of the ionization frequency was guaranteed by the non-equilibrium nature of the electron energy distribution function. In the direct current discharge, the tail of the electron distribution was populated by the high-energy electrons due to the passage of strata. In the radio-frequency discharge, the ratio between the electron energy relaxation time and driving voltage period was such that the distribution function does not react on the oscillating electric field but is defined by the effective electric field.

**DATA AVAILABILITY**

The data that support the finding of this study are available from the author upon reasonable request.

[1] P. S. Landa, N. A. Miskinova and Yu. V. Ponomarev, Sov. Phys. Usp. **23,** 813 (1980).

[2] V. I. Kolobov, J. Phys. D: Appl. Phys. **39,** R487 (2006).

[3] V. Desangles, J.-L. Raimbault, A Poye, P. Chabert, N. Plihon, Phys. Rev. Lett. **123,** 265001 (2019).

[4] Y. B. Golubovskii, V. O. Nekuchaev, A. Y. Skoblo, Tech. Phys. **59,** 1787 (2014).

[5] H. C. J. Mulders, W. J. M. Brok, W. W. Stoffels, IEEE Trans. Plasma Sci. **36,** 1380 (2008).

[6] Y.-X. Liu, E. Schüngel, I. Korolov, Z. Donko, Y.-N. Wang and J. Schulze, Phys. Rev. Lett. **116,** 255002 (2016).

[7] R. R. Arslanbekov and V. I. Kolobov, Phys. Plasmas **26,** 104501 (2019).

[8] V. I. Kolobov, R. R. Arslanbekov, D. Levko and V. A. Godyak, J. Phys. D: Appl. Phys. **53,** 25LT01 (2020).

[9] Yu. B. Golubovskii, V. O. Nekuchaev, A. Yu. Skoblo, Technical Physics **59,** 1787 (2014).

[10] Yu. B. Golubovskii, A. A. Kudryavtsev, V. O. Nekuchaev, I. A. Porokhova, and L. D. Tsendin, Electron Kinetic in Non-Equilibrium Gas Discharge Plasma (Saint Petersburg, 2004) (in Russian).

[11] F. Iza, S. S. Yang, H. C. Kim and J. K. Lee, J. Appl. Phys. **98,** 043302 (2005).

[12] E. Kawamura, M. A. Lieberman, A. Lichtenberg, Phys. Plasmas **25,** 013535 (2018).

[13] D. Levko and L. L. Raja, J. Appl. Phys. **121,** 093302 (2017).

[14] T. Ruzicka and K. Rohlena, Czechoslovak Journal Phys. B **22,** 906 (1972).

[15] Yu. B. Golubovskii, V. I. Kolobov, and V. O. Nekuchaev, Phys. Plasmas **20,** 101602 (2013).

[16] Yu. B. Golubovskii, V. A. Maiorov, I. A. Porokhova, and J. Behnke, J. Phys. D: Appl. Phys. **32,** 1391 (1999).

[17] L. D. Tsendin, Plasma Source Sci. Technol. **4,** 200 (1995).

[18] Yu. B. Golubovskii, S. Valin, E. Pelyukhova, V. Nekuchaev, and F. Sigeneger, Phys. Plasmas **23,** 123518 (2016).

[19] D. Levko, Phys. Plasmas **27,** 023505 (2020).

[20] Biagi's database for the electron-argon collision cross sections: www.lxcat.net/Biagi, retrieved on Sept.,15, 2020

[21] G. J. M. Hagelaar, Modeling methods for low-temperature plasmas, Habilitation à Diriger des Recherches, 2009.

[22] Yu. P. Raizer, Gas Discharge Physics (Dolgoprudnii, Intellect, 2009) (in Russian).

[23] D. Levko and L. L. Raja, Plasma Research Express **1,** 035005 (2019).

[24] Yu. B. Golubovskii, A. Yu. Skoblo, C. Wilke, R. V. Kozakov, and V. O. Nekuchaev, Plasma Source Sci. Technol. **18,** 045022 (2009).

[25] G. Bekefi, J. L. Hirshfield, and S. C. Brown, Phys. Fluids **4,** 173 (1961).

[26] C. Yuan, J. Yao, E. A. Bogdanov, A. A. Kudryavtsev, and Z. Zhou, Phys. Rev. E **101,** 031202 (2020).

[27] G. J. M. Hagelaar, L. C. Pitchford, Plasma Sources Sci. Technol. **14,** 722 (2005).